\documentclass[aps,prl,twocolumn,superscriptaddress]{revtex4-1}
\usepackage{graphicx}  
\usepackage{dcolumn}   
\usepackage{bm}        
\usepackage{amssymb}   
\usepackage{hyperref}
\usepackage{color}

\begin{document}

\title{Electric Dipole Polarizability of $^{48}$Ca and Implications for the Neutron Skin}

\newcommand{\TUDarmstadt}{Institut f\"ur Kernphysik, Technische Universit\"{a}t Darmstadt, 64289 Darmstadt, Germany}
\newcommand{\TRIUMF}{TRIUMF, 4004Wesbrook Mall, Vancouver, British Columbia V6T 2A3, Canada}
\newcommand{\UBC}{Department of Physics and Astronomy, University of British Columbia, Vancouver, British Columbia V6T 1Z4, Canada}
\newcommand{\UManitoba}{Department of Physics and Astronomy, University of Manitoba,Winnipeg, Manitoba R3T 2N2, Canada}
\newcommand{\TexasAM}{Department of Physics and Astronomy, Texas A\&M University-Commerce, Commerce, Texas 75429, USA}
\newcommand{\OakRidge}{Physics Division, Oak Ridge National Laboratory, Oak Ridge, Tennessee 37831, USA}
\newcommand{\UTennessee}{Department of Physics and Astronomy, University of Tennessee, Knoxville, Tennessee 37996, USA}
\newcommand{\RCNP}{Research Center for Nuclear Physics, Osaka University, Ibaraki, Osaka 567-0047, Japan}
\newcommand{\TWMU}{Tokyo Women's Medical University, 8-1 Kawada-cho, Shinjuku-ku, Tokyo 162-8666, Japan}
\newcommand{\EMMI}{ExtreMe Matter Institute EMMI, GSI Helmholtzzentrum f\"ur Schwerionenforschung GmbH, 64291 Darmstadt, Germany}
\newcommand{\MPIK}{Max-Planck-Institut f\"ur Kernphysik, Saupfercheckweg 1, 69117 Heidelberg, Germany}

\author{J.~Birkhan}\affiliation{\TUDarmstadt}
\author{M.~Miorelli}\affiliation{\TRIUMF}\affiliation{\UBC}
\author{S.~Bacca}\affiliation{\TRIUMF}\affiliation{\UManitoba}
\author{S.~Bassauer}\affiliation{\TUDarmstadt}
\author{C.~A.~Bertulani}\affiliation{\TexasAM}
\author{G.~Hagen}\affiliation{\OakRidge}\affiliation{\UTennessee}
\author{H.~Matsubara}\affiliation{\RCNP}\affiliation{\TWMU}
\author{P.~von~Neumann-Cosel}\email[Email:]{vnc@ikp.tu-darmstadt.de}\affiliation{\TUDarmstadt}
\author{T.~Papenbrock}\affiliation{\OakRidge}\affiliation{\UTennessee}
\author{N.~Pietralla}\affiliation{\TUDarmstadt}
\author{V.~Yu.~Ponomarev}\affiliation{\TUDarmstadt}
\author{A.~Richter}\affiliation{\TUDarmstadt}
\author{A.~Schwenk}\affiliation{\TUDarmstadt}\affiliation{\EMMI}\affiliation{\MPIK}
\author{A.~Tamii}\affiliation{\RCNP}

\begin{abstract}

The electric dipole strength distribution in $^{48}$Ca between 5 and
25~MeV has been determined at RCNP, Osaka, from proton inelastic
scattering experiments at forward angles.  Combined with
photoabsorption data at higher excitation energy, this enables the first extraction of the electric dipole polarizability
$\alpha_\mathrm{D}(^{48}{\rm Ca}) = 2.07(22)$~fm$^3$. Remarkably, the
dipole response of $^{48}$Ca is found to be very similar to that of
$^{40}$Ca, consistent with a small neutron skin in $^{48}$Ca.  The
experimental results are in good agreement with {\it ab initio}
calculations based on chiral effective field theory interactions and
with state-of-the-art density-functional calculations, implying a
neutron skin in $^{48}$Ca of $0.14 - 0.20$~fm.

\end{abstract}

\date{\today}

\maketitle 

{\em Introduction.}-- The equation of state (EOS) of neutron-rich
matter governs the properties of neutron-rich nuclei, the structure of
neutron stars, and the dynamics of core-collapse
supernovae~\cite{Lattimer2012,Hebeler2015}. The largest uncertainty of
the EOS at nuclear densities for neutron-rich conditions stems from
the limited knowledge of the symmetry energy $J$, which
is the difference of the energies of neutron and nuclear matter at
saturation density, and the slope of the symmetry energy $L$, which is
related to the pressure of neutron matter.

The symmetry energy also plays an important role in nuclei, where it
contributes to the formation of neutron skins in the presence of a neutron
excess.  Calculations based on energy density functionals (EDFs)
pointed out that $J$ and $L$ can be correlated with isovector collective
excitations of the nucleus such as pygmy dipole resonances~\cite{car10} and giant dipole resonances
(GDRs)~\cite{Trippa2008}, thus suggesting that the neutron skin
thickness, the difference of the neutron and proton root-mean-square
radii, could be constrained by studying properties of collective
isovector observables at low energy~\cite{kra99}. One such
observable is the nuclear electric dipole polarizability $\alpha_D$,
which represents a viable tool to constrain the EOS of neutron matter
and the physics of neutron
stars~\cite{brown2000,furnstahl2002,Tsang2012,ApJ2013,Hebeler2014,Brow14skyrme}.

While correlations among $\alpha_D$, the neutron skin and the symmetry
energy parameters have been studied extensively with
EDFs~\cite{rei10,roc13,roc15,Colo2015,Mondal2016}, only recently have
\textit{ab initio} calculations based on chiral effective field theory
($\chi$EFT) interactions successfully studied such correlations in
medium-mass nuclei~\cite{hag16,Miorelli2016}. By using a set of chiral
two- plus three-nucleon interactions~\cite{Hebeler11,Ekstroem15} and
exploiting correlations between $\alpha_D$ and the proton and neutron
radii, \citeauthor{hag16} predicted for the first time the electric
dipole polarizability and a neutron skin thickness of $0.12 - 0.15$~fm
for $^{48}$Ca from an \textit{ab initio}
calculation~\cite{hag16}. Since the electric dipole polarizability can
be measured rather precisely, this offers novel insights into the
properties of neutron-rich matter from a study of the dipole response
of $^{48}$Ca. The properties of neutron-rich matter also connect this
to the physics of the neutron-rich calcium isotopes, with recent
pioneering measurements of the masses and $2^+$ excitation energies up
to $^{54}$Ca~\cite{Wien13Nat,Step13Ca54} and of the charge radius up
to $^{52}$Ca~\cite{Ruiz16Calcium}.

The neutron skin thickness can be obtained by comparison of matter
radii deduced, e.g., from elastic proton scattering~\cite{sta94,zen10}
or coherent photoproduction of neutral pions~\cite{tar14} with
well-known charge radii from elastic electron scattering.  It can also
be measured directly with antiproton
annihilation~\cite{klo07,bro07}. A direct determination of the neutron
radius is possible with parity-violating elastic electron
scattering. Such an experiment (PREX) has been perfomed at JLAB for
$^{208}$Pb but at present statistical uncertainties limit the
precision~\cite{abr12}. An further measurement is approved and a
similar experiment on $^{48}$Ca (CREX) is presently under
discussion~\cite{rio13,hor14}. Here, we focus on the electric dipole
polarizability,
\begin{equation}\label{eq:dp}
\alpha_D = \frac{8 \pi}{9}\int\frac{B({\rm E1},E_{\rm X})}{E_{\rm X}} dE_{\rm X}  = \frac{\hbar c}{2 \pi^2} \int \frac{\sigma_\gamma(E_{\rm X})}{E_{\rm X}^2} dE_{\rm X} \,,
\end{equation}
where $B$(E1) and $\sigma_\gamma$ denote the electric dipole (E1)
strength distribution and the E1 photoabsorption cross section,
respectively, and $E_{\rm X}$ is the excitation energy.  The
evaluation of Eq.~(\ref{eq:dp}) requires a measurement of the complete
E1 strength distribution which is dominated by the GDR~\cite{ber75}.

A promising new method to measure the E1 strength distribution from
low energies across the GDR is inelastic proton scattering under
extreme forward angles including $0^\circ$ at energies of a few
hundred MeV~\cite{tam09,nev11}.  In these kinematics the cross
sections are dominated by relativistic Coulomb excitation, while the
nuclear excitation of collective modes, except for the spinflip $M1$
resonance~\cite{hey10}, is suppressed.  Results for $\alpha_D$
extracted for $^{208}$Pb~\cite{tam11} and $^{120}$Sn~\cite{has15} have been
shown to provide important constraints~\cite{tam14}  on the respective neutron
skins of these nuclei and, together with data on the exotic nucleus
$^{68}$Ni from experiments in inverse kinematics~\cite{ros13}, on
EDFs~\cite{roc15}. In this Letter, we report a measurement for
the electric dipole polarizability of $^{48}$Ca, which provides the first opportunity to compare with results from
{\it ab initio} calculations based on $\chi$EFT interactions and with
state-of-the-art EDF calculations in the same nucleus.
The insight gained will also impact on the proposed CREX experiment.

{\em Experiments}.-- The $^{48}$Ca$(p,p')$ reaction has been measured
at RCNP, Osaka, with an incident proton energy of 295 MeV.  Data were
taken with the Grand Raiden spectrometer~\cite{fuj99} in the
laboratory scattering angle range $0^\circ - 5.5^\circ$ for excitation
energies $5 - 25$ MeV.  A $^{48}$Ca foil with an isotopic enrichment
of 95.2\% and an areal density of 1.87 mg/cm$^2$ was bombarded with
proton beams of 4 to 10 nA.  Dispersion matching techniques
were applied to achieve an energy resolution of about 25 keV (full
width at half maximum).  The experimental techniques and the raw data
analysis are described in Ref.~\cite{tam09} while details for the
present experiment can be found in Ref.~\cite{bir15}.  

\begin{figure}[t]
\includegraphics[width=8.6cm]{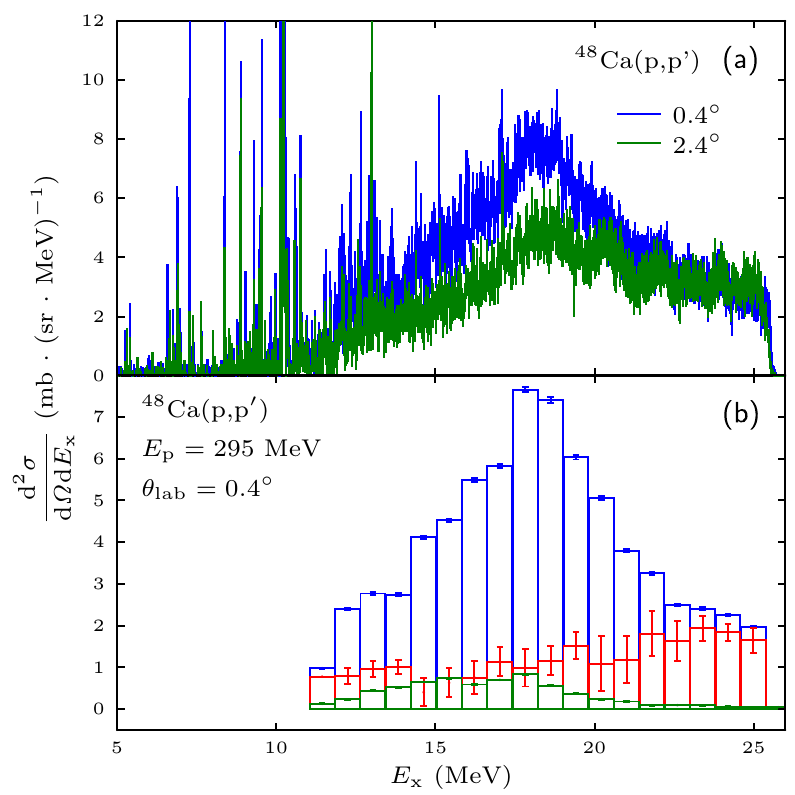}
\caption{(Color online)
(a) Spectra of the $^{48}$Ca$(p,p')$ reaction at $E_0 = 295$~MeV and scattering angles $\Theta_{\rm lab} = 0.4^\circ$ and $2.4^\circ$.
(b) Example of the decomposition for the spectrum at  $\Theta_{\rm lab} = 0.4^\circ$. 
Green histogram: Contribution from isoscalar giant resonances subtracted prior to the MDA.
Blue histogram: E1 part from the MDA.
Red histogram: Nuclear backgorund from the MDA.   
\label{Ca48-DP-fig1}}
\end{figure}

Figure~\ref{Ca48-DP-fig1}(a) shows representative spectra taken at
laboratory scattering angles $\Theta_{\rm lab} = 0.4^\circ$ (blue) and $2.4^\circ$. 
At lower excitation energies, a few discrete transitions are observed, mostly of E1
character~\cite{bir15}.  The prominent transition at 10.23 MeV has M1 character~\cite{bir16}. 
The cross sections above 10 MeV
show a broad resonance structure identified with excitation of the
GDR.  The decrease of cross sections with increasing scattering
angle is consistent with relativistic Coulomb excitation.

Cross sections due to relativistic Coulomb excitation can be separated from the
spinflip M1 resonance dominating the nuclear response at small
momentum transfers using spin transfer observables~\cite{tam11,has15}
or a multipole decomposition analysis (MDA) of angular
distributions~\cite{pol12,kru15}.  Comparison of the two independent
methods shows good agreement.  No polarization measurements
were performed for $^{48}$Ca since about 75\% of the spinflip M1 strength is concentrated in the transition at 10.23 MeV, while the rest is strongly fragmented into about 30 transitions between 7 and 13 MeV\cite{mat17}.

An angle-independent nuclear background due to quasifree scattering \cite{hau91} was included in the MDA.  
An example of the resulting MDA decomposition is presented in Fig.~\ref{Ca48-DP-fig1}(b).
In order to reduce the degrees of freedom in the $\chi^2$ minimization procedure, the cross
sections from excitation of the isoscalar giant monopole and
quadrupole resonance were determined from the experimental strength
functions in $^{48}$Ca~\cite{liu11} with the method described in
Ref.~\cite{kru15} and subtracted from the spectra.  The contributions
to the cross sections shown as green histogram in Fig.~\ref{Ca48-DP-fig1}(b) are small at the most forward angle (below 10\% in any given energy bin). 

\begin{figure}[t]
\includegraphics[width=8.6cm]{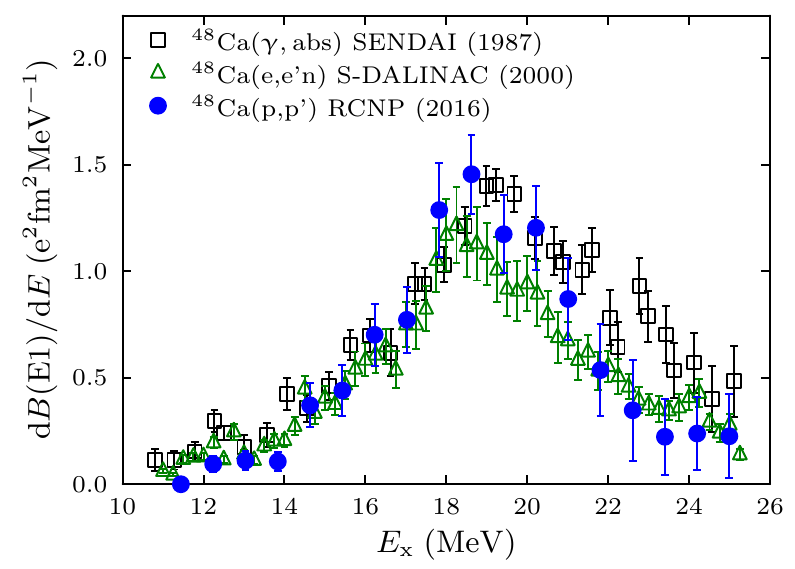}
\caption{(Color online)
Comparison of $B$(E1) strength distributions in $^{48}$Ca deduced from Ref.~\cite{oke87} (squares), Ref.~\cite{str00} (triangles), and from the present work (circles).
\label{Ca48-DP-fig2}}
\end{figure}

{\em E1 strength and photoabsorption cross sections}.--The Coulomb excitation cross sections resulting from the MDA were converted into equivalent photoabsorption cross sections and
a $B$(E1) strength distribution, respectively, using the virtual photon
method~\cite{ber88}.  The virtual photon spectrum was calculated in an
eikonal approach~\cite{ber93}.
The resulting $B$(E1) strength distribution is displayed as full
circles in Fig.~\ref{Ca48-DP-fig2}. The error bars include systematic
uncertainties of the absolute cross sections due to charge collection,
dead time of the data acquisition, target thickness, as well as a
variation of the minimum impact parameter in the calculation of the
virtual photon spectrum. The model dependence of the MDA was
considered by including the variance of $\chi^2$ values obtained for
fits with all possible combinations of theoretical input curves.  The
latter contribution dominates the overall uncertainty.

There exist two other measurements of E1 strength in $^{48}$Ca in the
energy region of the GDR.  A form factor decomposition of a
$^{48}$Ca($e,e^\prime n)$ experiment at the S-DALINAC~\cite{str00}
resulted in the strength distribution shown as open triangles in
Fig.~\ref{Ca48-DP-fig2}.  Considering that the error bars shown do not
include an additional 10\% uncertainty from the model dependence of
the form factor decomposition~\cite{str00} the two data sets are in
good agreement.  However, the proton emission contributes to the cross
sections above threshold ($S_p = 15.8$ MeV) although it is expected to
be weak in a neutron-rich nucleus. Another result~\cite{oke87} (open
squares) shows rather large deviations at the high-energy flank of
the GDR. It was obtained from excitation functions of the activity of
residual isotopes after particle emission. The photoabsorption cross
sections were deduced in an unfolding procedure with the
bremsstrahlung spectrum as input~\cite{pen59} leading to sizable
systematic uncertainties not reflected in the quoted error bars.
Furthermore, the contribution from the $(\gamma,2n)$ channel
contributing at higher $E_{\rm X}$ was estimated from statistical
model calculations assuming a large fraction of direct decay
inconsistent with the results of Ref.~\cite{str00}.  Thus, these
results are discarded in the following discussion.

\begin{figure}[t]
\includegraphics[width=8.6cm]{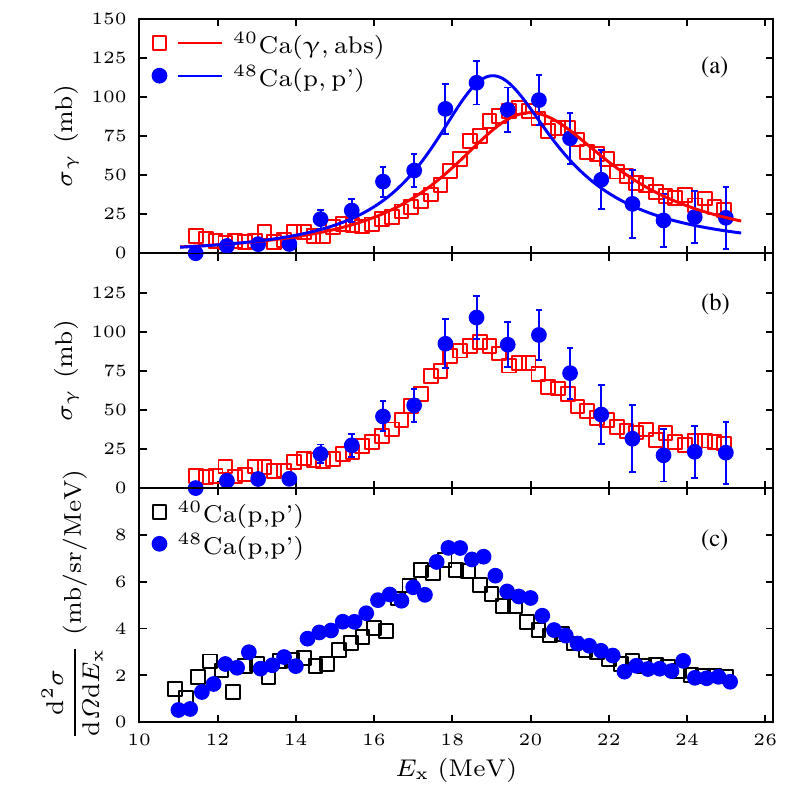}
\caption{(Color online)
(a) Photoabsorption cross sections in $^{48}$Ca (present work, circles) compared with $^{40}$Ca (Ref.~\cite{ahr75,ahr85},  squares).
(b) $^{40}$Ca data shifted by $-0.87$~MeV (Eq.~\ref{eq:gdrec}).
(c) Cross sections of the $(p,p')$ reaction at $E_0 = 295$~MeV and scattering angle $\Theta_{\rm lab} = 0.4^\circ$ for $^{48}$Ca (circles) and $^{40}$Ca (squares).
\label{Ca48-DP-fig3}}
\end{figure}
From the present work, photoabsorption cross sections in the range
$E_{\rm X}= 10 - 25$~MeV could be extracted and are displayed in
Fig.~\ref{Ca48-DP-fig3}(a) as solid dots. They are well described
by a Lorentzian with a centroid energy $E_{\rm C} = 18.9(2)$~MeV and a
width $\Gamma = 3.9(4)$~MeV.  The centroid energy is consistent with
systematics of the mass dependence~\cite{har01}
\begin{equation}
E_{\rm C} = 31.2 \, A^{-1/3} + 20.6 \, A^{-1/6}.
\label{eq:gdrec}
\end{equation}
The integrated strength in the measured energy range corresponds to an
exhaustion of the E1 energy-weighted sum rule of 85\%.  It is
instructive to compare to photoabsorption data for $^{40}$Ca (open
squares)~\cite{ahr85} which again are well described by a Lorentzian.
Figure \ref{Ca48-DP-fig3}(b) compares the two data sets after shifting
the $^{40}$Ca centroid by the amount predicted by
Eq.~(\ref{eq:gdrec}). It is evident that the GDR in $^{40}$Ca and
$^{48}$Ca exhibit nearly identical widths.  The contributions to the
electric dipole polarizability from the energy region $10 - 25$~MeV
are $\alpha_D(^{40}{\rm Ca}) = 1.50(2)$~fm$^3$ and $\alpha_D(^{48}{\rm
  Ca}) = 1.73(18)$~fm$^3$.

\begin{figure}[t]
\includegraphics[width=8.6cm]{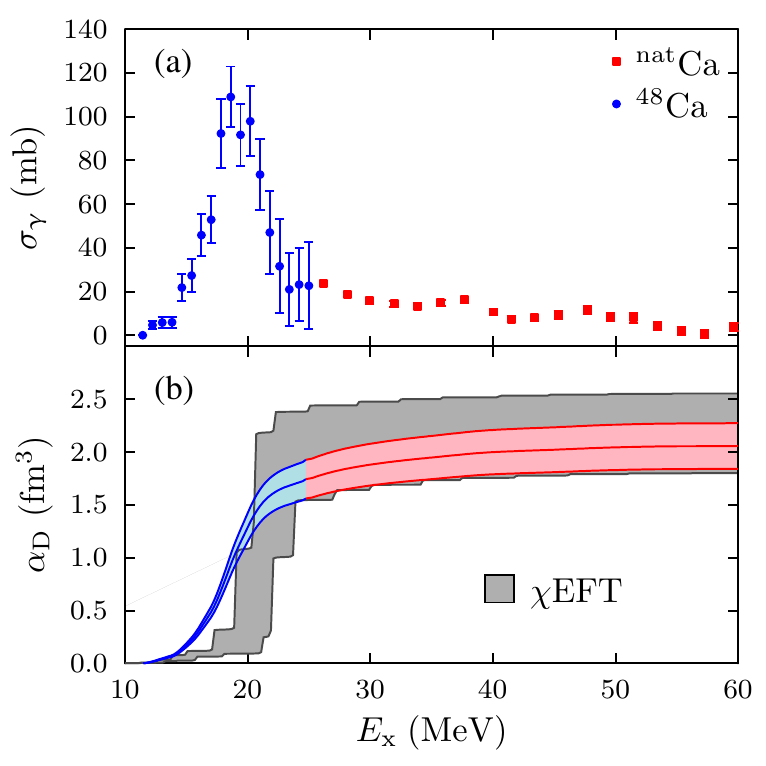}
\caption{(Color online) (a) Combined photoabsorption cross sections in
$^{48}$Ca from the present work (blue circles) for $E_{\rm X} \leq
25$~MeV and from Ref.~\cite{ahr75} (red squares) for $25 \leq E_{\rm
X} \leq 60$~MeV.  (b) Running sum of the electric dipole
polarizability in comparison to $\chi$EFT predictions, where the gray
band is based on a set of two- plus three-nucleon
interactions~\cite{hag16} and includes a partial uncertainty estimate from
the many-body method.
\label{Ca48-DP-fig4}}
\end{figure}

Although the GDR strength dominates, contributions to
$\alpha_D(^{48}{\rm Ca})$ at lower and higher excitation energies must
be considered.  Electric dipole strength below the neutron threshold
($S_n = 9.9$~MeV) was measured with the $(\gamma,\gamma^\prime)$
reaction~\cite{har02}.  Unlike in heavy nuclei, where the low-energy
strength is a sizable correction~\cite{pol12,kru15}, the contribution
$[0.0101(6)$~ fm$^3]$ is negligibly small in $^{48}$Ca. For the energy
region above 25~MeV, in analogy to the procedure described in
Ref.~\cite{has15} we adopt the $^{40}$Ca photoabsorption data of
Ref.~\cite{ahr75}, but shifted by the difference of centroid energies
for mass-48 and 40 predicted by
Eq.~(\ref{eq:gdrec}). Figure~\ref{Ca48-DP-fig4}(a) summarizes the
combined data used for the determination of $\alpha_D(^{48}{\rm Ca})$.

The data in Ref.~\cite{ahr75} extend up to the pion threshold
energy. However, here we evaluate $\alpha_D$ integrating the strength
up to 60~MeV since, as will be shown in the following paragraphs, the
sum rule is already well converged at these energies.
With these assumptions we deduce $\alpha_D(^{48}{\rm Ca}) =
2.07(22)$~fm$^3$.

For the comparison with theory it is instructive to also extract a
corresponding value for $^{40}$Ca, which one would expect to be
smaller than the one for $^{48}$Ca. As shown in Ref.~\cite{Bacca2014},
integrating the data for $^{40}$Ca from Ref.~\cite{ahr75} one obtains
$\alpha_D(^{40}{\rm Ca}) = 1.95(26)$~fm$^3$. Here, we combine the data
of Ref.~\cite{ahr75} with a refined set of data in the giant resonance
region measured by the same group~\cite{ahr85} and find
$\alpha_D(^{40}{\rm Ca}) = 1.87(3)$~fm$^3$. We note that a much higher
value was quoted in Ref.~\cite{ahr75} which would actually exceed our
result for $^{48}$Ca. The preference of the data set from
Ref.~\cite{ahr85} is motivated by a preliminary comparison with
$^{40}$Ca$(p,p^\prime)$ results taken at Osaka.  Although no $E1$
strength has been extracted yet, a comparison of spectra at the most
forward angles [Fig.~\ref{Ca48-DP-fig3}(c)], again shifted by the
centroid energy difference, demonstrates good correspondence of the
Coulomb excitation cross sections and an absolute ratio similar to the
one observed in Fig.~\ref{Ca48-DP-fig3}(b).

{\it Comparison with theory}.-- First principles calculations of
$\sigma_\gamma(E_{\rm x})$ require the solution of the many-body
scattering problem at all energies $E_{\rm x}$, including those in the
continuum, which is extremely challenging beyond few-nucleon
systems. While an \textit{ab initio} calculation of the full continuum
is still out of reach for medium-mass nuclei, methods based on
integral transforms that avoid its explicit
computation~\cite{Efros85,Efros94,Efros07} have been successfully
applied to light nuclei (see Ref.~\cite{BaccaPastore2014} for a
review) and recently extended to medium-mass
nuclei~\cite{Bacca13,Bacca2014,Hagen2014} using coupled-cluster
theory. Furthermore, it has been shown that energy-dependent sum
rules, such as the polarizability, can be evaluated without the
explicit knowledge of the continuum states or a cross-section
calculation itself~\cite{LSR} and recent
developments~\cite{Miorelli2016} have also allowed the calculation of
$\alpha_D$ as a function of the upper integration limit of
Eq.~(\ref{eq:dp}).

We performed \textit{ab initio} calculations of $\alpha_D$ using the
Lorentz integral transform coupled-cluster method described in
Refs.~\cite{Bacca2014,Miorelli2016}. The theoretical results are
compared to experiment in Fig.~\ref{Ca48-DP-fig4}(b), where the smooth
band (blue and red) shows the running sum of the experimental dipole
polarizability with error bars. The laddered (gray) band is based on
different chiral Hamiltonians, using the same two- and three-nucleon
interactions employed in Ref.~\cite{hag16}, which reproduce well
saturation properties of nuclear
matter~\cite{Ekstroem15,Hebeler11,Drischler2016}. For each
interaction, the estimated model-space dependence and truncation
uncertainty is about $4\%$ of $\alpha_D$, which is also included in
the gray band. We find that the agreement between the experimental and
theoretical results in Fig.~\ref{Ca48-DP-fig4}(b) is better for higher
excitation energies. However, we also observed that the position of
the GDR is more affected by truncations, which could lead to a shift
of $\approx 2\, \rm{MeV}$. In addition, we estimated that the
contributions from coupled-cluster triples corrections (due to genuine three-particle-three-hole correlations) could be
important at low energies. Both of these truncation errors are not
included in the uncertainty shown in Fig.~\ref{Ca48-DP-fig4}(b),
because it is difficult to quantify them without explicit
calculations. With these taken into account, the steep rise in the
theoretical band around $20\,\rm{MeV}$ indicates the position of the
GDR peak is consistent with the experimental centroid.

\begin{figure}[t]
\includegraphics[width=8.6cm]{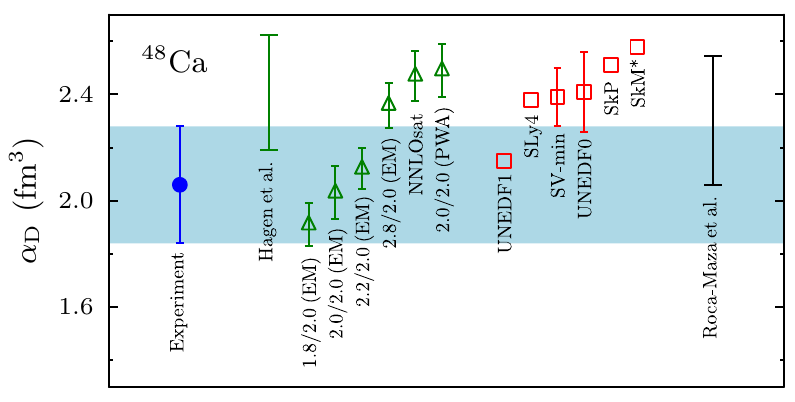}
\caption{(Color online)
Experimental electric dipole polarizability in $^{48}$Ca
(blue band) and predictions from $\chi$EFT (green triangles) and EDFs
(red squares, for details on the functionals see~\cite{hag16}, error bars from Ref.~\cite{WitekPrivate}).  The
green and black bar indicate the $\chi$EFT prediction selected to
reproduce the $^{48}$Ca charge radius~\cite{hag16} and the range of $\alpha_D$ predictions~\cite{roc15} from EDFs 
providing a consistent description of polarizabilities in $^{68}$Ni~\cite{ros13}, $^{120}$Sn~\cite{has15}, and $^{208}$Pb~\cite{tam11}, respectively.
\label{Ca48-DP-fig5}}
\end{figure}
In Fig.~\ref{Ca48-DP-fig5}, we present a detailed comparison of the
experimental $\alpha_D$ value with predictions from $\chi$EFT and
state-of-the-art EDFs. For the $\chi$EFT predictions (green triangles)
are based on a set of chiral two- plus three-nucleon interactions~\cite{Hebeler11,Ekstroem15}
whereas the EDF results are based the functionals SkM$^*$, SkP, SLy4, SV-min, 
UNEDF0 and UNEDF1~\cite{hag16}. In addition, we show a $\chi$EFT prediction selected to
reproduce the $^{48}$Ca charge radius~\cite{hag16} and the range of $\alpha_D$ predictions~\cite{roc15} from EDFs 
providing a consistent description of polarizabilities in $^{68}$Ni~\cite{ros13}, $^{120}$Sn~\cite{has15}, and $^{208}$Pb~\cite{tam11}. Taking only the interactions and functionals in Fig.~\ref{Ca48-DP-fig5}
consistent with the experimental range implies a neutron skin in
$^{48}$Ca of $0.14 - 0.20$~fm, where the lower neutron skin in this range ($< 0.15$~fm) is given
by the \textit{ab initio} calculations~\cite{hag16}.
For the latter, the small neutron skin is related to the strong $N=28$ shell closure, which leads to practically the same charge radii for $^{40}$Ca and $^{48}$Ca.

The {\em ab initio} results also provide symmetry energy parameter ranges $J = 28.5 - 33.3$ MeV and $L =  43.8 - 48.6$ MeV.
These constraints are highly competitive, in particular the value of $L$,  as can be seen in a current comparison of constraints from different methods \cite{lat16}.  
The EDF results show larger scattering, in particular for the density dependence \cite{roc15}.


{\em Summary}.-- We presented the first determination of the electric
dipole polarizability of $^{48}$Ca using relativistic Coulomb
excitation in the $(p,p^\prime)$ reaction at very forward angles. The
resulting dipole response of $^{48}$Ca is found to be remarkably
similar to that of $^{40}$Ca, consistent with a small neutron skin in
$^{48}$Ca. The result is in good agreement with predictions from
$\chi$EFT and EDF calculations pointing to a neutron skin of
$0.14-0.20$~fm. 

\begin{acknowledgments}
We thank W.\ Nazarewicz and X.\ Roca-Maza for useful discussions, and J. Lynn for a critical reading of the manuscript. This
work was supported by the DFG, 
Grant No.\ SFB 1245, JSPS KAKENHI Grant No.\ JP14740154, MEXT KAKENHI
Grant No.\ JP25105509, NSERC Grant No.\ SAPIN-2015-00031, the US DOE
Grants No.\ DE-FG02-08ER41533 (Texas A\&M University-Commerce),
DE-FG02-96ER40963, DE-SC0008499 (NUCLEI SciDAC collaboration), and the Field Work Proposal ERKBP57 at Oak Ridge National Laboratory (ORNL).
TRIUMF receives federal funding via a contribution agreement with the
National Research Council of Canada. 
Computer time was provided by the
INCITE program. This research used computing resources at TRIUMF and
of the Oak Ridge Leadership Computing Facility supported by the US DOE
under Contract No.\ DE-AC05-00OR22725.

This manuscript has been authored by UT-Battelle, LLC under
  Contract No. DE-AC05-00OR22725 with the U.S. Department of
  Energy. The United States Government retains and the publisher, by
  accepting the article for publication, acknowledges that the United
  States Government retains a non-exclusive, paid-up, irrevocable,
  world-wide license to publish or reproduce the published form of
  this manuscript, or allow others to do so, for United States
  Government purposes. The Department of Energy will provide public
  access to these results of federally sponsored research in
  accordance with the DOE Public Access
  Plan. (http://energy.gov/downloads/doe-public-access-plan).
\end{acknowledgments}

\bibliography{refs_total}

\end{document}